\documentclass[showpacs,aps,showkeys,graphicx,twocolumn,]{revtex4}
\usepackage{amsmath}
\usepackage{mathrsfs}
\usepackage{amsfonts}
\usepackage{amssymb}
\usepackage{graphicx}
\usepackage{eufrak}
\usepackage{multirow}
\usepackage{longtable}
\usepackage{float}
\usepackage{xcolor}
\usepackage{booktabs}

\usepackage{soul,color,xcolor}
\usepackage[colorlinks,linkcolor=blue,anchorcolor=blue,citecolor=blue]{hyperref}

\begin{document}

\title{Remote state preparation of single-partite high-dimensional states in complex Hilbert spaces}

\author{Jun-Hai Zhao$^1$, Si-Qi Du$^1$, Wen-Qiang Liu$^2$, Dong-Hong Zhao$^{1}$ and Hai-Rui Wei$^{1*}$}
\email{hrwei@ustb.edu.cn}
\address{$^1$ School of Mathematics and Physics, University of Science and Technology Beijing, Beijing 100083, China}

\address{$^2$ Department of Mathematics and Physics, Shijiazhuang Tiedao University, Shijiazhuang 050043}

\date{\today }

\begin{abstract}
High-dimensional quantum systems offer a new playground for quantum information applications due to their remarkable advantages such as higher capacity and noise resistance.
We propose potentially practical schemes for remotely preparing four- and eight-level equatorial states in complex Hilbert spaces exactly by identifying a set of orthogonal measurement bases.
In these minimal-resource-consuming schemes, both pre-shared maximally and non-maximally entangled states are taken into account.
The three-, five-, six-, and seven-level equatorial states in complex Hilbert spaces can also be obtained by adjusting the parameters of the desired states.
The evaluations indicate that our high-dimensional RSP schemes might be possible with current technology.
The collection operations, necessary for our  high-dimensional RSP schemes via partially entangled channels, can be avoided by encoding the computational basis in the spatial modes of single-photon systems.
	\end{abstract}



\maketitle

\section{Introduction}\label{sec1}

Remote state preparation (RSP) \cite{Pati,Lo,Bennett} is one of the remarkable protocols for remotely preparing a quantum state without transmitting the quantum system physically.
RSP is sometimes called teleportation of a known state.
Both the quantum teleportation (QT)  \cite{Teleportation1,Teleportation2,Teleportation3} and RSP are used for transmitting a quantum state from a sender to a long-distance receiver with usage of a previously shared maximally entangled state \cite{entanglement1,entanglement2,entanglement3,entanglement4} and some classical information without physically sending the particles.
In QT, the teleported state is owned by the sender, the information of the teleported state is unknown to both sender and receiver, and one maximally entangled state (ebit) and two bits of classical communication (cbits) are both necessary and sufficient.
Unlike usual QT, in RSP, the information of the state to be prepared is known fully to sender but unknown to receiver, in particular, sender need not own the state.
Note that for the constrained state, RSP is possible to trade off between ebit and cbits \cite{Lo,Bennett,faithful-RSP} which renders RSP  more economical than QT.
Besides, RSP may be preferable to avoid QT in full Bell-type measurements, which currently an outstanding experimental challenge resorting to linear optics only \cite{BSA}.

The first RSP was proposed for certain special qubits on the Bloch sphere in 2001 \cite{Pati,Lo}.
Since then, many theoretical and experimental improvements have been proposed \cite{improved}, such as
low-entanglement RSP  \cite{low-entanglement-RSP},
high-dimensional RSP \cite{Zeng,qudit-Song},
generalized RSP \cite{generalized-RSP},
oblivious RSP \cite{oblivious-RSP},
continuous variable RSP \cite{continuous-variable-RSP2},
mixed state RSP \cite{optimal-RSP1,mixed3},
optimal RSP \cite{optimal-RSP1},
faithful RSP \cite{faithful-RSP},
bidirectional RSP \cite{bidirectional},
joint RSP \cite{joit},
multi-degree-of-freedom RSP \cite{two-degree-of-freedom RSP},
and cyclic RSP \cite{cyclic}.
Recently, increasing effort has been devoted toward high-dimensional RSP \cite{qudit-Song} due to the remarkable features of high-dimensional systems, for example, increased information capacity \cite{capacity2,Chang_capacity},
stronger violation of Bell's nonlocality \cite{Bell-non-locality2},
simplified experimental setups  \cite{photon2-efficiency},
improved computing efficiency \cite{photon2-efficiency},
better security in quantum communication \cite{better-security-noise-resistance1},
and enhanced noise-resistance \cite{better-security-noise-resistance3}.

Despite  the advantages of high-dimensional quantum systems, they are rarely investigated both theoretically and experimentally.
Nowadays, high-dimensional quantum systems have been benefited wide range of applications, such as
circuit building \cite{high-circuit1,high-circuit2,high-circuit3,high-circuit4,high-circuit5},
entanglement generation \cite{high-entanglement1,high-entanglement2,high-entanglement3,high-entanglement4,high-entanglement5,high-entanglement6},
quantum algorithm design \cite{high-algorithm},
quantum state generation \cite{state-generation1,state-generation2},
entanglement purification \cite{purify},
quantum key distribution (QKD) \cite{QKD1,QKD2},
entanglement distribution \cite{distribution},
high-dimensional measurement \cite{measurement},
and Bell test \cite{Bell-test}.
For high-dimensional RSP, early in 2001, Zeng and Zhang \cite{Zeng} proved that RSP can only be implemented in two-, four- or eight-dimensional real Hilbert subspace.
Later in 2006, Yu \emph{et al}. \cite{qudit-Song} proposed a technique for remotely preparing equatorial states with real coefficients using maximally entangled states of qubits.
In 2022, Ma \emph{et al}. \cite{cyclic} presented a cyclic controlled RSP in three-dimensional system.
However, high-dimensional RSP are mainly focused on  the real subspace within the complex Hilbert space, and architectures for RSP in high-dimensional complex Hilbert spaces have, so far, been absent.

In this paper, we present protocols to remotely prepare four- and eight-dimensional equatorial states in complex Hilbert spaces.
We first propose a RSP in four-dimensional complex Hilbert space using a maximally entangled pair.
This maximally-entangled-based scheme is accomplished by exploiting single-partite Von Neumann measurement and some single-partite operations.
Then, we generalize the scheme to the non-maximally entangled channel case.
Subsequently, we extend above RSP schemes to eight-dimensional complex Hilbert space case, and the nontrivial orthogonal bases is presented in detail.
Lastly, we evaluate the performances of our high-dimensional RSP schemes.

Our high-dimensional RSP schemes have the following characteristics. Our schemes are all implemented by identifying a set of orthogonal measurement basis, and this projection measurement is easier to implement in physical experiments than POVM measurements.
Compared to the cluster-based RSP \cite{cyclic}, our RSP minimizes the number of entanglement pairs.
In addition, the three-dimensional (five-, six-, and seven-dimensional) RSP can also be obtained by adjusting the parameter of the our four-dimensional (eight-dimensional) minimum RSP. It is worthy to point out here that if  we encode the qunits ($s$-level with $s= 4,\;8$) in the spatial mode states of single-photon systems,
the necessary $s$-dimensional collective operation can be exactly achieved by employing $(s-1)$ variable reflectivity beam splitters (VBSs).

The rest of the paper is organized as follows.  In Sec. \ref{sec2},  we present RSPs in four-dimensional complex Hilbert space.
Here the pre-shared maximally and non-maximally entangled states are taken into account, respectively.
In Sec. \ref{sec3}, we generalize the protocols to the eight-dimensional Hilbert space cases.
The performances of our high-dimensional RSP schemes are evaluated in Sec. \ref{sec4}.
In Sec. \ref{sec5}, we provide discussions and conclusions.

\section{Remote preparation of single-qudit state} \label{sec2}

\subsection{Four-dimensional RSP using maximally entangled state as quantum channel}\label{sec21}

Let us consider a single-qudit (4-level) equatorial state
\begin{equation}\label{eq1}
|\psi_0\rangle= c_0|0\rangle + c_1 e^{\text{i}\theta_1} |1\rangle + c_2 e^{\text{i}\theta_2} |2\rangle + c_3 e^{\text{i}\theta_3} |3\rangle,
\end{equation}
where real coefficient $c_i$ ($i=0,1,2,3$) satisfy normalization condition $\sum_{i=0}^3 c_i^2=1$, and real parameter $\theta_i\in[0,\;2\pi)$.
The information about $|\psi_0\rangle$ is only known by Alice.

Now  Alice (the sender) wants to remotely prepare $|\psi_0\rangle$ at  Bob's (the receiver) site.
To reach this aim, firstly, Alice and Bob need to pre-share a maximally entangled state $|\Psi\rangle_{AB}$. Here
\begin{equation}\label{eq2}
|\Psi\rangle_{AB} = \frac{1}{2}(|00\rangle + |11\rangle + |22\rangle + |33\rangle)_{AB}.
\end{equation}
The subscripts $A$ ($B$) refer to the particle $A$ ($B$) possessed by Alice (Bob).

Subsequently, Alice performs a transformation
\begin{equation}\label{eq3}
U=\text{diag}\{1,e^{\text{i}2\theta_1},e^{\text{i}2\theta_2}, e^{\text{i}2\theta_3}\},
\end{equation}
on particle $A$. And then, $|\Psi\rangle_{AB}$ becomes
\begin{equation}\label{eq4}
|\Psi'\rangle_{AB} = \frac{1}{2}(|00\rangle + e^{\text{i}2\theta_1}|11\rangle + e^{\text{i}2\theta_2}|22\rangle + e^{\text{i}2\theta_3}|33\rangle)_{AB}.
\end{equation}
We rewrite $|\Psi'\rangle_{AB}$ in the orthogonal basis $\{|\psi_0\rangle, |\psi_1\rangle,|\psi_2\rangle,|\psi_3\rangle\}$, i.e.,
\begin{equation}\label{eq5}
\begin{split}
|\Psi'\rangle_{AB} =
  \frac{1}{2}(|\psi_0\rangle_A \otimes |\psi_0\rangle_B +|\psi_1\rangle_A \otimes |\psi_1\rangle_B\\
             +|\psi_2\rangle_A \otimes |\psi_2\rangle_B +|\psi_3\rangle_A \otimes |\psi_3\rangle_B).
\end{split}
\end{equation}
Here
\begin{equation}\label{eq6}
|\psi_0\rangle = + c_0|0\rangle  + c_1 e^{\text{i}\theta_1}|1\rangle  + c_2 e^{\text{i}\theta_2}|2\rangle  + c_3 e^{\text{i}\theta_3}|3\rangle,
\end{equation}
\begin{equation}\label{eq7}
|\psi_1\rangle = -c_1|0\rangle  + c_0 e^{\text{i}\theta_1}|1\rangle  - c_3 e^{\text{i}\theta_2}|2\rangle  + c_2 e^{\text{i}\theta_3}|3\rangle,
\end{equation}
\begin{equation}\label{eq8}
|\psi_2\rangle = -c_2|0\rangle  + c_3 e^{\text{i}\theta_1}|1\rangle  + c_0 e^{\text{i}\theta_2}|2\rangle  - c_1 e^{\text{i}\theta_3}|3\rangle,
\end{equation}
\begin{equation}\label{eq9}
|\psi_3\rangle = -c_3|0\rangle  - c_2 e^{\text{i}\theta_1}|1\rangle  + c_1 e^{\text{i}\theta_2}|2\rangle  + c_0 e^{\text{i}\theta_3}|3\rangle.
\end{equation}
That is, $|\psi_i\rangle = \text{diag} \{1, e^{\text{i}\theta_1}, e^{\text{i}\theta_2}, e^{\text{i}\theta_3}\}
\cdot \mathcal{T}_4 |i\rangle $, $i=0,1,2,3$. Here
\begin{equation}\label{eq10}
\mathcal{T}_4 = \left(
        \begin{array}{cccc}
          c_0 & -c_1 & -c_2 & -c_3 \\
          c_1 &  c_0 &  c_3 & -c_2 \\
          c_2 & -c_3 &  c_0 & c_1 \\
          c_3 &  c_2 & -c_1 & c_0\\
        \end{array}
      \right).
\end{equation}
Note that single-particle operation $\text{diag} \{1, e^{\text{i}\theta_1}, e^{\text{i}\theta_2}, e^{\text{i}\theta_3}\}$ is  relatively ease to implement. Here and henceforth, the implementation of $\text{diag} \{1, e^{\text{i}\theta_1}, e^{\text{i}\theta_2}, e^{\text{i}\theta_3}\}$ will not be considered further.

Nextly, Alice measures her particle $A$ in the basis $\{|\psi_0\rangle, |\psi_1\rangle,|\psi_2\rangle,|\psi_3\rangle\}$, and informs  Bob of her measurement result via a classical communication (one cdit).
Based on Eq. \eqref{eq5}, one can see that if the result of the measurement is $|\psi_0\rangle_A$, then leaving Bob in the desired state $|\psi_0\rangle_B$ with a success probability of 1/4.
If the result of the measurement is $|\psi_1\rangle_A$, $|\psi_2\rangle_A$, or $|\psi_3\rangle_A$, the RSP fails. This is because Bob cannot correct $|\psi_1\rangle_B, |\psi_2\rangle_B$, or $|\psi_3\rangle_B$ to $|\psi_0\rangle_B$ as the phase parameters $\theta_i$ of the target state $|\psi_0\rangle$ are unknown to Bob.

Note that if $\theta_1$, $\theta_2$ and $\theta_3$ given in Eq. \eqref{eq1} are restricted to $\theta_1=\theta_2=\theta_3=0$,
the four-dimensional RSP using maximally entangled state as quantum channel can be completed exactly. That because Bob can correct $|\psi_j\rangle_B$ ($j=1,\;2,\;3$) to $|\psi_0\rangle_B$ by applying some classical single-qudit feed-forward unitary operations $V_{j}$ on particle $B$.
In the basis $\{|0\rangle, |1\rangle, |2\rangle, |3\rangle \}$, operation $V_j$ can be expressed as
\begin{equation} \label{eq11}
  V_1=\left(
        \begin{array}{cccc}
          0 &  1  &  0  &  0   \\
         -1 &  0  &  0  &  0   \\
          0 &  0  &  0  &  1   \\
          0 &  0  & -1  &  0   \\
        \end{array}
      \right),
 \end{equation}
\begin{equation} \label{eq12}
  V_2 = \left(
         \begin{array}{cccc}
           0 &  0  &  1   &  0   \\
           0 &  0  &  0   & -1   \\
          -1 &  0  &  0   &  0   \\
           0 &  1  &  0   &  0   \\
         \end{array}
       \right),
\end{equation}
\begin{equation} \label{eq13}
 V_3 = \left(
         \begin{array}{cccc}
           0 & 0  &  0  &  1   \\
           0 & 0  &  1  &  0   \\
           0 &-1  &  0  &  0   \\
          -1 & 0  &  0  &  0   \\
         \end{array}
       \right).
  \end{equation}
Therefore, the success probability of the presented maximally-entangled-based RSP in four-dimensional {real subspace within the complex Hilbert space} can be improved to 100\% in principle.

\subsection{Four-dimensional RSP using non-maximally entangled state as quantum channel} \label{sec22}

Most of the quantum information processing tasks in fact work only with the maximally entangled states.
However, in practical applications, the entanglement will be inevitably degraded to another form due to the decoherence induced by its environment.
This degradation will decrease the fidelity and security of quantum communication.

In this subsection, the pre-shared normalization less-entangled pure state $|\tilde{\Psi}\rangle_{AB}$, instead of the maximally entangled state $|\Psi\rangle_{AB}$ given in Eq. \eqref{eq2}, is taken into account. Here
\begin{equation}\label{eq14}
|\tilde{\Psi}\rangle_{AB} = (a_0|00\rangle + a_1|11\rangle + a_2|22\rangle + a_3|33\rangle)_{AB}.
\end{equation}
Real coefficients $a_i$ is only known by Bob, and $|a_0|>|a_1|>|a_2|>|a_3|$.

A similar arrangement as that made in Sec. \ref{sec21}, after single-partite transformation $U$ given by Eq. \eqref{eq3} is applied on particle $A$ by Alice, $|\tilde{\Psi}\rangle_{AB}$ becomes
\begin{equation}\label{eq15}
	\begin{split}
|\tilde{\Psi}'\rangle_{AB} =\;&
(|\psi_0\rangle_A \otimes |\tilde{\psi}_0\rangle_B
             +|\psi_1\rangle_A \otimes |\tilde{\psi}_1\rangle_B\\&
             +|\psi_2\rangle_A \otimes |\tilde{\psi}_2\rangle_B
             +|\psi_3\rangle_A \otimes |\tilde{\psi}_3\rangle_B).
	\end{split}
\end{equation}
Here
\begin{equation}\label{eq16}
	\begin{split}
|\tilde{\psi}_0\rangle_B =\;& +a_0c_0|0\rangle  + a_1c_1 e^{\text{i}\theta_1}|1\rangle + a_2c_2 e^{\text{i}\theta_2}|2\rangle \\& + a_3c_3 e^{\text{i}\theta_3}|3\rangle,
	\end{split}
\end{equation}
\begin{equation}\label{eq17}
	\begin{split}
|\tilde{\psi}_1\rangle_B =\;& -a_0c_1|0\rangle  + a_1c_0 e^{\text{i}\theta_1}|1\rangle  - a_2c_3 e^{\text{i}\theta_2}|2\rangle  \\&+ a_3c_2 e^{\text{i}\theta_3}|3\rangle,
	\end{split}
\end{equation}
\begin{equation}\label{eq18}
	\begin{split}
|\tilde{\psi}_2\rangle_B =\;& -a_0c_2|0\rangle  + a_1c_3 e^{\text{i}\theta_1}|1\rangle + a_2c_0 e^{\text{i}\theta_2}|2\rangle  \\&- a_3c_1 e^{\text{i}\theta_3}|3\rangle,
	\end{split}
\end{equation}
\begin{equation}\label{eq19}
	\begin{split}
|\tilde{\psi}_3\rangle_B =\;& -a_0c_3|0\rangle  - a_1c_2 e^{\text{i}\theta_1}|1\rangle  + a_2c_1 e^{\text{i}\theta_2}|2\rangle  \\&+ a_3c_0 e^{\text{i}\theta_3}|3\rangle.
	\end{split}
\end{equation}

If the measurement result of Alice is $|\psi_0\rangle_A$, $|\tilde{\Psi}'\rangle_{AB}$ will collapse to $|\tilde{\psi}_0\rangle_B$.
In order to get $|\psi_0\rangle_B$,  Bob first introduces an auxiliary qudit $a$ with the original state $|0\rangle_a$.
Subsequently, Bob performs a 2-qudit collective unitary transformation $W$ on particles $B$ and $a$. In the basis
$\{|00\rangle_{aB}$,
  $|01\rangle_{aB}$,
  $|02\rangle_{aB}$,
  $|03\rangle_{aB}$,
  $|10\rangle_{aB}$,
  $|11\rangle_{aB}$,
  $|12\rangle_{aB}$,
  $|13\rangle_{aB}$,
  $|20\rangle_{aB}$,
  $|21\rangle_{aB}$,
  $|22\rangle_{aB}$,
  $|23\rangle_{aB}$,
  $|30\rangle_{aB}$,
  $|31\rangle_{aB}$,
  $|32\rangle_{aB}$,
  $|33\rangle_{aB}\}$,
operation $W$ can be expressed as
\begin{equation}\label{eq20}
W=
\left(
  \begin{array}{ccc}
    w_{11} & w_{12} & 0 \\
    w_{21} & w_{22} & 0 \\
      0    &   0    & I_8 \\
  \end{array}
\right),
\end{equation}
where $I_8$ is the $8 \times 8$ identity matrix, and
\begin{equation}\label{eq21}
w_{11}=-w_{22}=\text{diag}\{ \frac{a_3}{a_0},\frac{a_3}{a_1},\frac{a_3}{a_2},\frac{a_3}{a_3}\},
\end{equation}
\begin{equation}\label{eq22}
	\begin{split}
w_{12}= w_{21}= \text{diag} \{ &\sqrt{1-(\frac{a_3}{a_0})^2},
                                \sqrt{1 -(\frac{a_3}{a_1})^2}, \\
                               &\sqrt{1-(\frac{a_3}{a_2})^2}, 0\}.
\end{split}
\end{equation}
This transformation will convert $|\tilde{\psi}_0\rangle_B$ into the following state
\begin{equation}\label{eq23}
\begin{split}
|\tilde{\psi}_0'\rangle_{Ba} =\;& a_3(c_0|0\rangle+
                                  c_1 e^{\text{i}\theta_1}|1\rangle +
                                  c_2 e^{\text{i}\theta_2}|2\rangle  +
                                  c_3 e^{\text{i}\theta_3}|3\rangle)_B|0\rangle_a \\& +
                                 (c_0 \sqrt{(a_0)^2 - (a_3)^2} |0\rangle_{B}\\& +
                                  c_1 e^{\text{i}\theta_1} \sqrt{(a_1)^2 - (a_3)^2}|1\rangle_{B}\\& +
                                  c_2 e^{\text{i}\theta_2} \sqrt{(a_2)^2 - (a_3)^2}|2\rangle_{B})|1\rangle_{a}.
\end{split}
\end{equation}

Nextly, Bob makes a measurement on the auxiliary particle $a$ in the basis $\{|0\rangle_a, |1\rangle_a, |2\rangle_a, |3\rangle_a\}$.

If the result of the measurement is $|0\rangle_a$,
the $|\tilde{\psi}'_0\rangle_{Ba}$ will be projected into the desired state $|\psi_0\rangle_B$ with a success probability of $|a_3|^2$.
Fortunately, if we restrict to $\theta_1=\theta_2=\theta_3=0$, the success probability can be boosted to $4|a_3|^2$ in principle.
That is because, based on the result of the measurement $|\psi_j\rangle_A$ ($j=0,\;1,\;2,\;3$)  made by Alice, Bob can also convert into $|\tilde{\psi}_j\rangle_B$ to $|\tilde{\psi}_0\rangle_B$ by applying single-qudit classical feed-forward unitary operations $V_{j}$, given by Eq. \eqref{eq11} - Eq. \eqref{eq13}, on particle $B$.

If the result of the measurement is $|1\rangle_a$,  the state in $|\tilde{\psi}'_0\rangle_{Ba}$ will collapse to a less-size state
\begin{equation}\label{eq24}
\begin{split}
|\psi_{\text{small}}\rangle_{B} =\;&
                                  c_0 \sqrt{(a_0)^2 - (a_3)^2} |0\rangle_{B}\\& +
                                  c_1 e^{\text{i}\theta_1} \sqrt{(a_1)^2 - (a_3)^2}|1\rangle_{B}\\& +
                                  c_2 e^{\text{i}\theta_2} \sqrt{(a_2)^2 - (a_3)^2}|2\rangle_{B}.
\end{split}
\end{equation}
Obviously, the state given in Eq. \eqref{eq24} is not the desired state $|\psi_0\rangle_B$ given in Eq. \eqref{eq1}, that is to say, the scheme is failure. But it dose not mean the resulting less-size state given Eq. \eqref{eq24} is useless.

Based on Sec. \ref{sec21} and Sec. \ref{sec22}, we can see that the success probability and the efficiency of the high-dimension RSP scheme via non-maximally entangled state is lower than the counterpart via maximally entangled state. The quantum resources that are required to implement RSP via non-maximally entangled state are more than the counterpart via maximally entangled state. Moreover, the implementation of 2-qudit collective operation $W$ described in  Eq. \eqref{eq20} is challenging.

\section{Remote preparation of single-qunit (8-level) state}\label{sec3}

\subsection{Eight-dimensional RSP using maximally entangled state as quantum channel} \label{sec31}

The single-qunit (here 8-level is considered) normalization equatorial state which Alice wants Bob to prepared is given by
\begin{equation}\label{eq25}
\begin{split}
|\phi_0\rangle=\; &
             d_0|0\rangle
             + d_1 e^{\text{i}\vartheta_1} |1\rangle
             + d_2 e^{\text{i}\vartheta_2} |2\rangle
             + d_3 e^{\text{i}\vartheta_3} |3\rangle\\&
             + d_4 e^{\text{i}\vartheta_4} |4\rangle
             + d_5 e^{\text{i}\vartheta_5} |5\rangle
             + d_6 e^{\text{i}\vartheta_6} |6\rangle\\&
             + d_7 e^{\text{i}\vartheta_7} |7\rangle,
\end{split}
\end{equation}
where real parameters $d_i$ and $\vartheta_j$ are only known to Alice.
The maximally entangled state shared with Alice and Bob to complete above purpose can be written as
\begin{equation}\label{eq26}
\begin{split}
|\Phi\rangle_{AB} = \;&\frac{1}{2\sqrt{2}}(|00\rangle + |11\rangle + |22\rangle + |33\rangle +|44\rangle \\& +|55\rangle + |66\rangle + |77\rangle)_{AB}.
\end{split}
\end{equation}

In order to implement RSP in eight complex Hilbert space, Alice first performs an unitary transformation $X$ on particle $A$. Here $X$ is given by
\begin{equation}\label{eq27}
\begin{split}
X= \text{diag}\{& 1, e^{\text{i}2\vartheta_1},
                     e^{\text{i}2\vartheta_2},
                     e^{\text{i}2\vartheta_3},
                     e^{\text{i}2\vartheta_4},
                     e^{\text{i}2\vartheta_5},\\&
                     e^{\text{i}2\vartheta_6},
                     e^{\text{i}2\vartheta_7}\},
\end{split}
\end{equation}
And then $|\Phi\rangle_{AB}$  will be transformed into
\begin{equation}\label{eq28}
\begin{split}
|\Phi'\rangle_{AB}=\;&\frac{1}{2\sqrt{2}}(|00\rangle + e^{\text{i}2\vartheta_1}|11\rangle
                                                     + e^{\text{i}2\vartheta_2}|22\rangle\\&
                                                     + e^{\text{i}2\vartheta_3}|33\rangle
                                                     + e^{\text{i}2\vartheta_4}|44\rangle
                                                     + e^{\text{i}2\vartheta_5}|55\rangle\\&
                                                     + e^{\text{i}2\vartheta_6}|66\rangle
                                                     + e^{\text{i}2\vartheta_7}|77\rangle)_{AB}.
\end{split}
\end{equation}
In a set of new orthogonal vectors $\{|\phi_0\rangle$, $|\phi_1\rangle$, $\cdots$, $|\phi_7\rangle\}$, state  $|\Phi'\rangle_{AB}$ can be expanded as
\begin{equation}\label{eq29}
  |\Phi '{\rangle _{AB}} = \frac{1}{{2\sqrt 2 }}\sum\limits_{i = 0}^7 {|{\phi _i}{\rangle _A} \otimes |{\phi _i}{\rangle _B}}.
\end{equation}
  Here
\begin{equation}\label{eq30}
\begin{split}
|\phi _0\rangle =\;&+d_0 |0\rangle + d_1 e^{\text{i} \vartheta_1} |1\rangle
                                   + d_2 e^{\text{i} \vartheta_2} |2\rangle
                                   + d_3 e^{\text{i} \vartheta_3} |3\rangle\\&
                                   + d_4 e^{\text{i} \vartheta_4} |4\rangle
                                   + d_5 e^{\text{i} \vartheta_5} |5\rangle
                                   + d_6 e^{\text{i} \vartheta_6} |6\rangle\\&
                                   + d_7 e^{\text{i} \vartheta_7} |7\rangle,
\end{split}
\end{equation}
\begin{equation}\label{eq31}
\begin{split}
|\phi _1\rangle =\;& - d_1 |0\rangle + d_0 e^{\text{i} \vartheta_1} |1\rangle
                                     - d_3 e^{\text{i} \vartheta_2} |2\rangle
                                     + d_2 e^{\text{i} \vartheta_3} |3\rangle\\&
                                     - d_5 e^{\text{i} \vartheta_4} |4\rangle
                                     + d_4 e^{\text{i} \vartheta_5} |5\rangle
                                     + d_7 e^{\text{i} \vartheta_6} |6\rangle\\&
                                     - d_6 e^{\text{i} \vartheta_7} |7\rangle,
\end{split}
\end{equation}
\begin{equation}\label{eq32}
\begin{split}
|\phi _2\rangle =\;& - d_2 |0\rangle  + d_3 e^{\text{i} \vartheta_1} |1\rangle
                                      + d_0 e^{\text{i} \vartheta_2} |2\rangle
                                      - d_1 e^{\text{i} \vartheta_3} |3\rangle\\&
                                      - d_6 e^{\text{i} \vartheta_4} |4\rangle
                                      - d_7 e^{\text{i} \vartheta_5} |5\rangle
                                      + d_4 e^{\text{i} \vartheta_6} |6\rangle\\&
                                      + d_5 e^{\text{i} \vartheta_7} |7\rangle,
\end{split}
\end{equation}
\begin{equation}\label{eq33}
\begin{split}
|\phi _3\rangle =\;& - d_3 |0\rangle  - d_2 e^{\text{i} \vartheta_1} |1\rangle
                                      + d_1 e^{\text{i} \vartheta_2} |2\rangle
                                      + d_0 e^{\text{i} \vartheta_3} |3\rangle\\&
                                      - d_7 e^{\text{i} \vartheta_4} |4\rangle
                                      + d_6 e^{\text{i} \vartheta_5} |5\rangle
                                      - d_5 e^{\text{i} \vartheta_6} |6\rangle\\&
                                      + d_4 e^{\text{i} \vartheta_7} |7\rangle,
\end{split}
\end{equation}
\begin{equation}\label{eq34}
\begin{split}
|\phi _4\rangle =\;& - d_4 |0\rangle  + d_5 e^{\text{i} \vartheta_1} |1\rangle
                                      + d_6 e^{\text{i} \vartheta_2} |2\rangle
                                      + d_7 e^{\text{i} \vartheta_3} |3\rangle\\&
                                      + d_0 e^{\text{i} \vartheta_4} |4\rangle
                                      - d_1 e^{\text{i} \vartheta_5} |5\rangle
                                      - d_2 e^{\text{i} \vartheta_6} |6\rangle\\&
                                      - d_3 e^{\text{i} \vartheta_7} |7\rangle,
\end{split}
\end{equation}
\begin{equation}\label{eq35}
\begin{split}
|\phi _5\rangle =\;& - d_5 |0\rangle  - d_4 e^{\text{i} \vartheta_1} |1\rangle
                                      + d_7 e^{\text{i} \vartheta_2} |2\rangle
                                      - d_6 e^{\text{i} \vartheta_3} |3\rangle\\&
                                      + d_1 e^{\text{i} \vartheta_4} |4\rangle
                                      + d_0 e^{\text{i} \vartheta_5} |5\rangle
                                      + d_3 e^{\text{i} \vartheta_6} |6\rangle\\&
                                      - d_2 e^{\text{i} \vartheta_7} |7\rangle,
\end{split}
\end{equation}
\begin{equation}\label{eq36}
\begin{split}
|\phi_6\rangle =\;& - d_6 |0\rangle  - d_7 e^{\text{i} \vartheta_1} |1\rangle
                                     - d_4 e^{\text{i} \vartheta_2} |2\rangle
                                     + d_5 e^{\text{i} \vartheta_3} |3\rangle\\&
                                     + d_2 e^{\text{i} \vartheta_4} |4\rangle
                                     - d_3 e^{\text{i} \vartheta_5} |5\rangle
                                     + d_0 e^{\text{i} \vartheta_6} |6\rangle\\&
                                     + d_1 e^{\text{i} \vartheta_7} |7\rangle,
\end{split}
\end{equation}
\begin{equation}\label{eq37}
\begin{split}
|\phi _7\rangle =\; & - d_7 |0\rangle + d_6 e^{\text{i} \vartheta_1} |1\rangle
                                      - d_5 e^{\text{i} \vartheta_2} |2\rangle
                                      - d_4 e^{\text{i} \vartheta_3} |3\rangle\\&
                                      + d_3 e^{\text{i} \vartheta_4} |4\rangle
                                      + d_2 e^{\text{i} \vartheta_5} |5\rangle
                                      - d_1 e^{\text{i} \vartheta_6} |6\rangle\\&
                                      + d_0 e^{\text{i} \vartheta_7} |7\rangle.
\end{split}
\end{equation}
That is, $|\phi_j\rangle  = \text{diag}\{ 1, e^{\text{i}\vartheta_1},
                                             e^{\text{i}\vartheta_2}, \cdots,
                                             e^{\text{i}\vartheta_7}\} \cdot \mathcal{T}_8 |j\rangle$, $j=0,1,\cdots,7$.  Here
\begin{equation}\label{eq38}
\mathcal{T}_8 =
\left(
  \begin{array}{cccccccc}
    d_0 & -d_1 & -d_2 & -d_3 & -d_4 & -d_5 & -d_6 & -d_7 \\
    d_1 &  d_0 &  d_3 & -d_2 &  d_5 & -d_4 & -d_7 &  d_6 \\
    d_2 & -d_3 &  d_0 &  d_1 &  d_6 &  d_7 & -d_4 & -d_5 \\
    d_3 &  d_2 & -d_1 &  d_0 &  d_7 & -d_6 &  d_5 & -d_4 \\
    d_4 & -d_5 & -d_6 & -d_7 &  d_0 &  d_1 &  d_2 &  d_3 \\
    d_5 &  d_4 & -d_7 &  d_6 & -d_1 &  d_0 & -d_3 &  d_2 \\
    d_6 &  d_7 &  d_4 & -d_5 & -d_2 &  d_3 &  d_0 & -d_1 \\
    d_7 & -d_6 &  d_5 &  d_4 & -d_3 & -d_2 &  d_1 &  d_0 \\
  \end{array}
\right).
\end{equation}
Note that single-particle operation $\text{diag}\{ 1, e^{\text{i}\vartheta_1}, \cdots,  e^{\text{i}\vartheta_7}\}$ is much more easier to implement than $\mathcal{T}_8$.

Subsequently, Alice makes a single-partite measurement on particle $A$ in the basis $\{|\phi_0\rangle, |\phi_1\rangle,\cdots,|\phi_7\rangle\}$, and informs Bob of her measurement result via a classical communication (one cnit).
If Alice's measurement result is $|\phi_0\rangle_A$, then the state $|\Phi'\rangle_{AB}$ shown in Eq. \eqref{eq29} will collapse to the desired state $|\phi_0\rangle_B$ with a success probability of 1/8;
If the result of the measurement is $|\phi_j\rangle_A$ with $j=1,2,\cdots,7$, then the present RSP fails.

It is worthy of being noted that if we restrict to $\vartheta_1=\vartheta_2=\cdots=\vartheta_7=0$,
Bob can correct $|\phi_j\rangle_B$  ($j=1,2,\cdots,7$) to $|\phi_0\rangle_B$ by applying some classical single-qunit feed-forward unitary operation $Y_{j}$ on particle $B$.
In the basis $\{|0\rangle, |1\rangle, \cdots, |7\rangle \}$, single-qunit operation $Y_j$ can be expressed as
\begin{equation}  \label{eq39}
  Y_1= \left(
       \begin{array}{*{20}{cccc}}
         -\text{i} \sigma_y &        0            &        0          &  0         \\
               0            & -\text{i} \sigma_y  &        0          &  0         \\
               0            &      0              & -\text{i}\sigma_y &  0         \\
               0            &      0              &        0          &  \text{i}\sigma_y \\
        \end{array}
        \right),
\end{equation}
\begin{equation}  \label{eq40}
  Y_2 = \left(
        \begin{array}{*{20}{cccc}}
             0             & - \sigma_z     & 0           &   0  \\
         \sigma_z  &     0                   & 0          &   0  \\
             0            &     0                   & 0          & -I_2 \\
             0             &     0                  & I_2       &    0 \\
        \end{array}
        \right),
\end{equation}
\begin{equation}  \label{eq41}
  Y_3 = \left(
        \begin{array}{*{20}{cccc}}
         0            & -\sigma_x &       0                          &        0          \\
    \sigma_x &    0                &       0                          &        0          \\
         0           &    0              &       0                           & -\text{i}\sigma_y \\
         0          &    0                & -\text{i}\sigma_y &        0          \\
        \end{array}
        \right),
\end{equation}
\begin{equation}  \label{eq42}
   Y_4 =\left(
        \begin{array}{*{20}{cccc}}
            0           &   0     & -\sigma_z  & 0   \\
            0            &   0     &     0             & I_2 \\
        \sigma_z &   0      &     0             & 0   \\
            0            & -I_2 &     0              & 0   \\
        \end{array}
       \right),
\end{equation}
\begin{equation}  \label{eq43}
  Y_5 =\left(
       \begin{array}{*{20}{cccc}}
           0            &    0                             & -\sigma_x &       0   \\
           0           &    0                             &     0               & \text{i}\sigma_y\\
      \sigma_x &    0                             &     0               &       0   \\
          0             & \text{i}\sigma_y &     0               &       0    \\
       \end{array}
       \right),
\end{equation}
\begin{equation}  \label{eq44}
  Y_6 =\left(
       \begin{array}{*{20}{cccc}}
       0   &    0    &      0              & -I_2 \\
       0  &    0          & -\sigma_z  &   0  \\
       0  &\sigma_z &      0           &   0  \\
      I_2 &    0         &      0            &   0  \\
      \end{array}
      \right),
\end{equation}
\begin{equation}  \label{eq45}
  Y_7 =\left(
      \begin{array}{*{20}{cccc}}
           0     &     0    &      0     & -\sigma_y \\
           0     &     0    & -\sigma_x  &    0      \\
           0     & \sigma_x &      0     &    0      \\
      -i\sigma_y &     0    &      0     &    0       \\
       \end{array}
      \right),
\end{equation}
where
\begin{equation} \label{eq46}
\sigma_x = \left(
            \begin{array}{*{20}{cc}}
               0 & 1 \\
               1 & 0 \\
             \end{array}
            \right),\;
\sigma_y = \left(
            \begin{array}{*{20}{cc}}
              0         &  -\text{i} \\
              \text{i}  &      0     \\
              \end{array}
            \right),\;
\sigma_z = \left(
            \begin{array}{*{20}{cc}}
               1 & 0 \\
               0 & -1\\
            \end{array}
           \right).
\end{equation}
That is, the presented RSP in eight-dimensional  real subspace within the complex Hilbert space using maximally entangled state as quantum channel can be achieved in a deterministic way in principle.

\subsection{Eight-dimensional RSP using non-maximally entangled state as quantum channel} \label{sec32}

We consider the channel noise converts the maximally entangled state, given in Eq. \eqref{eq26}, into the following normalization { non-maximally entangled state}
\begin{equation}\label{eq47}
\begin{split}
|\tilde{\Phi}\rangle_{AB} =\;& (b_0|00\rangle + b_1|11\rangle + b_2|22\rangle + b_3|33\rangle \\
&+ b_4|44\rangle + b_5|55\rangle + b_6|66\rangle + b_7|77\rangle)_{AB}.
\end{split}
\end{equation}
Here real coefficient $b_i$ $(i=0,1,\cdots,7)$ is only known by Alice,  and $|b_0|>|b_1|>\cdots>|b_7|$.

Using the same procedure as  that made in Sec. \ref{sec31}, after Alice performs transformation $X$, given by Eq. \eqref{eq27}, on particle $A$, state $|\tilde{\Phi}\rangle_{AB}$ can be expanded as
\begin{equation}\label{eq48}
|\Phi '{\rangle _{AB}} = \;\sum\limits_{i = 0}^7 {|{\phi _i}{\rangle _A} \otimes |{{\tilde \phi }_i}{\rangle _B}}.
\end{equation}
Here
\begin{equation}\label{eq49}
\begin{split}
|\tilde{\phi}_0\rangle_B =\;&
+ d_0 b_0 | 0 \rangle
+ d_1 b_1 e^{\text{i} \vartheta_1} | 1 \rangle
+ d_2 b_2 e^{\text{i} \vartheta_2} | 2 \rangle\\&
+ d_3 b_3 e^{\text{i} \vartheta_3} | 3 \rangle
+ d_4 b_4 e^{\text{i} \vartheta_4} | 4 \rangle
+ d_5 b_5 e^{\text{i} \vartheta_5} | 5 \rangle\\&
+ d_6 b_6 e^{\text{i} \vartheta_6} | 6 \rangle
+ d_7 b_7 e^{\text{i} \vartheta_7} | 7 \rangle,
\end{split}
\end{equation}
\begin{equation}\label{eq50}
\begin{split}
|\tilde{\phi}_1\rangle_B =\;&
- d_1 b_0 | 0 \rangle
+ d_0 b_1 e^{i \vartheta_1} | 1 \rangle
- d_3 b_2 e^{i \vartheta_2} | 2 \rangle\\&
+ d_2 b_3 e^{i \vartheta_3} | 3 \rangle
- d_5 b_4 e^{i \vartheta_4} | 4 \rangle
+ d_4 b_5 e^{i \vartheta_5} | 5 \rangle\\&
+ d_7 b_6 e^{i \vartheta_6} | 6 \rangle
- d_6 b_7 e^{i \vartheta_7} | 7 \rangle,
\end{split}
\end{equation}
\begin{equation}\label{eq51}
\begin{split}
|\tilde{\phi}_2\rangle_B =\;&
- d_2 b_0 | 0 \rangle
+ d_3 b_1 e^{i \vartheta_1} | 1 \rangle
+ d_0 b_2 e^{i \vartheta_2} | 2 \rangle\\&
- d_1 b_3 e^{i \vartheta_3} | 3 \rangle
- d_6 b_4 e^{i \vartheta_4} | 4 \rangle
- d_7 b_5 e^{i \vartheta_5} | 5 \rangle\\&
+ d_4 b_6 e^{i \vartheta_6} | 6 \rangle
+ d_5 b_7 e^{i \vartheta_7} | 7 \rangle,
\end{split}
\end{equation}
\begin{equation}\label{eq52}
\begin{split}
|\tilde{\phi}_3\rangle_B =\;&
- d_3 b_0 | 0 \rangle
- d_2 b_1 e^{i \vartheta_1} | 1 \rangle
+ d_1 b_2 e^{i \vartheta_2} | 2 \rangle\\&
+ d_0 b_3 e^{i \vartheta_3} | 3 \rangle
- d_7 b_4 e^{i \vartheta_4} | 4 \rangle
+ d_6 b_5 e^{i \vartheta_5} | 5 \rangle\\&
- d_5 b_6 e^{i \vartheta_6} | 6 \rangle
+ d_4 b_7 e^{i \vartheta_7} | 7 \rangle,
\end{split}
\end{equation}
\begin{equation}\label{eq53}
\begin{split}
|\tilde{\phi}_4\rangle_B =\;&
- d_4 b_0 | 0 \rangle
+ d_5 b_1 e^{i \vartheta_1} | 1 \rangle
+ d_6 b_2 e^{i \vartheta_2} | 2 \rangle\\&
+ d_7 b_3 e^{i \vartheta_3} | 3 \rangle
+ d_0 b_4 e^{i \vartheta_4} | 4 \rangle
- d_1 b_5 e^{i \vartheta_5} | 5 \rangle\\&
- d_2 b_6 e^{i \vartheta_6} | 6 \rangle
- d_3 b_7 e^{i \vartheta_7} | 7 \rangle,
\end{split}
\end{equation}
\begin{equation}\label{eq54}
\begin{split}
|\tilde{\phi}_5\rangle_B =\;&
- d_5 b_0 | 0 \rangle
- d_4 b_1 e^{i \vartheta_1} | 1 \rangle
+ d_7 b_2 e^{i \vartheta_2} | 2 \rangle\\&
- d_6 b_3 e^{i \vartheta_3} | 3 \rangle
+ d_1 b_4 e^{i \vartheta_4} | 4 \rangle
+ d_0 b_5 e^{i \vartheta_5} | 5 \rangle\\&
+ d_3 b_6 e^{i \vartheta_6} | 6 \rangle
- d_2 b_7 e^{i \vartheta_7} | 7 \rangle,
\end{split}
\end{equation}
\begin{equation}\label{eq55}
\begin{split}
|\tilde{\phi}_6\rangle_B =\;&
- d_6 b_0 | 0 \rangle
- d_7 b_1 e^{i \vartheta_1} | 1 \rangle
- d_4 b_2 e^{i \vartheta_2} | 2 \rangle\\&
+ d_5 b_3 e^{i \vartheta_3} | 3 \rangle
+ d_2 b_4 e^{i \vartheta_4} | 4 \rangle
- d_3 b_5 e^{i \vartheta_5} | 5 \rangle\\&
+ d_0 b_6 e^{i \vartheta_6} | 6 \rangle
+ d_1 b_7 e^{i \vartheta_7} | 7 \rangle,
\end{split}
\end{equation}
\begin{equation}\label{eq56}
\begin{split}
|\tilde{\phi}_7\rangle_B =\; &
- d_7 b_0 | 0 \rangle
+ d_6 b_1 e^{i \vartheta_1} | 1 \rangle
- d_5 b_2 e^{i \vartheta_2} | 2 \rangle\\&
- d_4 b_3 e^{i \vartheta_3} | 3 \rangle
+ d_3 b_4 e^{i \vartheta_4} | 4 \rangle
+ d_2 b_5 e^{i \vartheta_5} | 5 \rangle\\&
- d_1 b_6 e^{i \vartheta_6} | 6 \rangle
+ d_0 b_7 e^{i \vartheta_7} | 7 \rangle.
\end{split}
\end{equation}

Subsequently, Alice performs single-partite projective measurement on her particle $A$ in the basis $\{|\phi_0\rangle, |\phi_1\rangle,\cdots,|\phi_7\rangle\}$, and informs Bob of her measurement outcome.

Based on Eq. \eqref{eq48}, we can find that if Alice's measurement result is $|\phi_0\rangle_A$, then $|\tilde{\Phi}'\rangle_{AB}$ will collapse to $|\tilde{\phi}_0\rangle_B$.
In order to concentrate the desired state $|\phi_0\rangle_B$, shown in Eq. \eqref{eq25}, from $|\tilde{\phi}_0\rangle_B$,  Bob  first borrows an auxiliary qunit $a$ with the original state $|0\rangle_a$, and then performs a 2-qunit collective operation $Z$ on particles $B$ and $a$.
The matrix representation of $Z$ in the basis
$\{|00\rangle_{aB}$,
  $|01\rangle_{aB}$,
  $|02\rangle_{aB}$,
  $|03\rangle_{aB}$,
  $|04\rangle_{aB}$,
  $|05\rangle_{aB}$,
  $|06\rangle_{aB}$,
  $|07\rangle_{aB}$,
  $\cdots$,
  $|70\rangle_{aB}$,
  $|71\rangle_{aB}$,
  $|72\rangle_{aB}$,
  $|73\rangle_{aB}$,
  $|74\rangle_{aB}$,
  $|75\rangle_{aB}$,
  $|76\rangle_{aB}$,
  $|77\rangle_{aB}\}$
is given by
\begin{equation}\label{eq57}
Z=
\left(
  \begin{array}{ccccccc}
    z_{11} & z_{12} & 0     \\
    z_{21} & z_{22} & 0      \\
      0          &   0           & I_{48} \\
  \end{array}
\right),
\end{equation}
where
\begin{equation}\label{eq58}
z_{11} = -z_{22}= \text{diag}\big\{\frac{b_7}{b_0},
                                   \frac{b_7}{b_1},
                                   \frac{b_7}{b_2},
                                   \frac{b_7}{b_3},
                                   \frac{b_7}{b_4},
                                   \frac{b_7}{b_5},
                                   \frac{b_7}{b_6},
                                   \frac{b_7}{b_7}\big\},
\end{equation}
\begin{equation}\label{eq59}
\begin{split}
z_{12}=z_{21}=\text{diag}\big\{&\sqrt{1 - (\frac{b_7}{b_0})^2},
                                \sqrt{1 - (\frac{b_7}{b_1})^2},\\&
                                \sqrt{1 - (\frac{b_7}{b_2})^2},
                                \sqrt{1 - (\frac{b_7}{b_3})^2},\\&
                                \sqrt{1 - (\frac{b_7}{b_4})^2},
                                \sqrt{1 - (\frac{b_7}{b_5})^2},\\&
                                \sqrt{1 - (\frac{b_7}{b_6})^2},0\big\}.
\end{split}
\end{equation}
After operation $Z$ is applied on particles $B$ and $a$, state $|\tilde{\phi}_0\rangle_B$ is transformed into
\begin{equation}\label{eq60}
\begin{split}
|\tilde{\phi}'_0\rangle_{Ba}=\;&
   b_7 (d_0 |0\rangle
       +d_1 e^{\rm{i}\vartheta_1}|1\rangle
       +d_2 e^{\rm{i}\vartheta_2}|2\rangle\\&
       +d_3 e^{\rm{i}\vartheta_3}|3\rangle
       +d_4 e^{\rm{i}\vartheta_4}|4\rangle
       +d_5 e^{\rm{i}\vartheta_5}|5\rangle\\&
       +d_6 e^{\rm{i}\vartheta_6}|6\rangle
       +d_7 e^{\rm{i}\vartheta_7}|7\rangle)_B |0\rangle_a \\&
+\big(d_0 \sqrt{1 - (\frac{b_7}{b_0})^2} |0\rangle\\&
     +d_1 e^{\rm{i}\vartheta_1} \sqrt{1 - (\frac{b_7}{b_1})^2} |1\rangle\\&
     +d_2 e^{\rm{i}\vartheta_2} \sqrt{1 - (\frac{b_7}{b_2})^2} |2\rangle\\&
     +d_3 e^{\rm{i}\vartheta_3} \sqrt{1 - (\frac{b_7}{b_3})^2} |3\rangle\\&
     +d_4 e^{\rm{i}\vartheta_4} \sqrt{1 - (\frac{b_7}{b_4})^2} |4\rangle\\&
     +d_5 e^{\rm{i}\vartheta_5} \sqrt{1 - (\frac{b_7}{b_5})^2} |5\rangle\\&
     +d_6 e^{\rm{i}\vartheta_6} \sqrt{1 - (\frac{b_7}{b_6})^2} |6\rangle \big)_B|1\rangle_a.
\end{split}
\end{equation}

Lastly, Bob measures auxiliary particle $a$ in the basis $\{|0\rangle_a, |1\rangle_a, \cdots, |7\rangle_a\}$.

If Bob's measurement result is $|0\rangle_a$, the system composed of particles $a$ and $B$ will be projected into the desired state
$|\phi_0\rangle_B$ with a success probability of $|b_7|^2$.
Note that if we restrict to $\vartheta_1=\cdots=\vartheta_7=0$, based on Alice's measurement result $|\phi_j\rangle_A $ $(j=2,\cdots,7)$, Bob can enhance the scheme success probability to $8|b_7|^2$ by performing $Y_j$, given in Eq. \eqref{eq39} - Eq. \eqref{eq45}, on state $|\tilde{\phi}_j\rangle_B$. This means
\begin{equation}\label{eq62}
Y_j|\tilde{\phi}_j\rangle_B  =|\tilde{\phi}_0\rangle_B, \;j=1,2,\cdots,7.
\end{equation}

If the result of the measurement is $|1\rangle_a$, the state $|\tilde{\phi}'_0\rangle_{Ba}$ in Eq. \eqref{eq60} will be projected into a less-size state
\begin{equation}\label{eq61}
\begin{split}
|\phi_{\text{small}}\rangle_{B} =\;&(d_0 \sqrt{1 - (\frac{b_7}{b_0})^2} |0\rangle\\&
                                    +d_1 e^{\rm{i}\vartheta_1} \sqrt{1 - (\frac{b_7}{b_1})^2} |1\rangle\\&
                                    +d_2 e^{\rm{i}\vartheta_2} \sqrt{1 - (\frac{b_7}{b_2})^2} |2\rangle\\&
                                    +d_3 e^{\rm{i}\vartheta_3} \sqrt{1 - (\frac{b_7}{b_3})^2} |3\rangle\\&
                                    +d_4 e^{\rm{i}\vartheta_4} \sqrt{1 - (\frac{b_7}{b_4})^2} |4\rangle\\&
                                    +d_5 e^{\rm{i}\vartheta_5} \sqrt{1 - (\frac{b_7}{b_5})^2} |5\rangle\\&
                                    +d_6 e^{\rm{i}\vartheta_6} \sqrt{1 - (\frac{b_7}{b_6})^2} |6\rangle)_B.
\end{split}
\end{equation}
This means the scheme fails.

\section{The evaluations of the high-dimensional RSP schemes}\label{sec4}

High-dimensional quantum systems, high-dimensional entangled states, single-partite unitary operations $\mathcal{T}_4$ given in Eq. \eqref{eq10} and $\mathcal{T}_8$ given in Eq. \eqref{eq38}, and bipartite collective operations $W$ given in Eq. \eqref{eq20} and $Z$ given in Eq. \eqref{eq57} are the crucial components of our proposals. In this section, we will evaluate the performances of these components.

\subsection{The evaluations of the high-dimensional quantum systems and the high-dimensional entangled states}\label{sec41}

High-dimensional systems have been implemented in various physical systems, including photons \cite{photon1,photon2,photon3,photon4}, ion traps \cite{ion}, nitrogen-vacancy centers \cite{NV}, nuclear magnetic resonances \cite{NMR1}, molecular magnets \cite{molecular}, and superconductors \cite{superconduct2}.
In those platforms, photon has been recognized as one of the most prominent natural candidates for qudit due to its many qudit-like orthogonal optical modes \cite{spatial1},
such as path, orbital angular momentum (OAM), time-bins, and  frequency.
Nowadays, high-dimensional entangled states in spatial (path) mode \cite{spatial1,spatial2}, OAM \cite{OAM1,OAM2,OAM3,OAM2017}, and energy-time \cite{energy-time1} have been reported.
Among all these degrees of freedom (DOFs), the spatial DOFs has excellent dimension scalability, a very high fidelity and is easy to manipulate.
Hence, in our schemes, the computational basis ${|j\rangle}$ is encoded in the path ($j=0,1,2,\ldots $) of single-photon systems.

\subsection{The evaluations of the transformations $\mathcal{T}_4$ and $\mathcal{T}_8$}\label{sec42}

Mutually unbiased basis \cite{MUB2002,MUB2013,chang2024MUB} are a crucial class of orthogonal measurement basis in high-dimensional systems, and it play an important role in quantum state characterization, cryptography, and entangled state certification.
Von Neumann measurement is the core component of RSP scheme for remotely preparing single-partite states.

\begin{figure}[htbp]
\centering
  \includegraphics[width=8.5 cm]{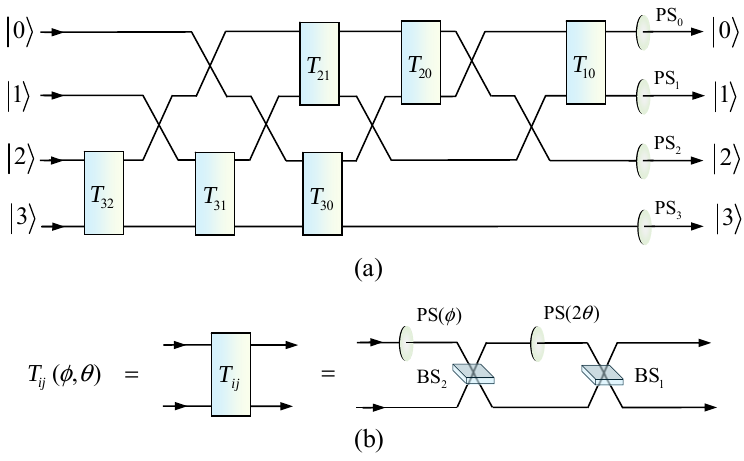}
  \caption{(a) Schematic setup to implement $\mathcal{T}_4$ with $c_0=c_1=c_2=c_3=\frac{1}{2}$. The phase shifts 0, $\pi$, $\pi$, $\pi$ are taken for  PS$_{0}$, PS$_{1}$, PS$_{2}$, PS$_{3}$, respectively.
  (b) Schematic setup to implement block $T(\phi,\theta)$ \cite{BS1994}.
  Phase shifter $\text{PS}(\theta)=e^{\text{i}\theta}$. BS is a 50:50 beam splitter.
  } \label{Figure1}
\end{figure}

It is known that Reck \cite{BS1994} and Clements \cite{BS2016} schemes show that arbitrary spatial transformations can be implemented by employing the Mach-Zehnder Interferometers (MZIs) and phase shifters (PSs). As shown in Fig.  \ref{Figure1}(b), standard MZI is composed of two 50:50 beam splitters (BSs) and one PS \cite{BS1994},  $U_{\text{MZI}}(2\theta)= U_{\text{BS}_1} \cdot U_{\text{PS}} (2\theta) \cdot U_{\text{BS}_2}$.
The matrix forms of the 50:50 BS, phase shifter $\text{PS}(2\theta)$, and MZI are given by
\begin{equation}\label{eq63}
U_{\text{BS}} = \frac{1}{\sqrt{2}}
    \left(
      \begin{array}{cc}
        1              &      \text{i} \\
        \text{i} &        1 \\
      \end{array}
    \right),
\end{equation}
\begin{equation}\label{eq64}
U_\text{PS}(2\theta) =
    \left(
      \begin{array}{cc}
        e^{\text{i}2\theta}      &     0 \\
             0                                  &       1 \\
      \end{array}
    \right),
\end{equation}
\begin{equation}\label{eq65}
\begin{split}
U_{\text{MZI}}(2\theta) = \text{i} e^{\text{i}\theta}
    \left(
       \begin{array}{cc}
            \sin\theta & \cos\theta \\
            \cos\theta & -\sin\theta \\
       \end{array}
     \right).
\end{split}
\end{equation}
The matrix form of the block $T(\phi,\theta)$ can be calculated as
\begin{equation}\label{eq66}
T(\phi,\theta) = \text{i} e^{\text{i}\theta}
\left(
  \begin{array}{cc}
    e^{\text{i}\phi}\sin \theta  &  \cos \theta \\
    e^{\text{i}\phi} \cos \theta & - \sin \theta \\
  \end{array}
\right).
\end{equation}
An overall phase $ \text{i} e^{\text{i}\theta}$ is omitting thereafter, since it is irrelevant for quantum information processing tasks.
Note that Eq. \eqref{eq66} is intrinsically the matrix form of VBS, and the reflectivity coefficient can be changed by adjusting $\phi$ and $\theta$.

Spatial single-qudit operations $\mathcal{T}_4$ given in Eq. \eqref{eq10} and $\mathcal{T}_8$ given in Eq. \eqref{eq38} can be implemented by employing sequences of $T_{\phi ,\theta}$ in principle.
For the sake of clarity, we take the scenario of $\mathcal{T}_4$ with $c_0=c_1=c_2=c_3=\frac{1}{2}$ as a representative example. As shown in Fig. \ref{Figure1}(a), $\mathcal{T}_4$ with $c_0=c_1=c_2=c_3=\frac{1}{2}$ can be decomposed as

\begin{equation}\label{eq67}
\mathcal{D} \cdot T_{10} \cdot T_{20} \cdot T_{21} \cdot T_{30} \cdot T_{31} \cdot T_{32},
\end{equation}
where
\begin{equation}\label{eq68}
\mathcal{D}  =\left(
     \begin{array}{cccc}
       1 &  0 &  0 &  0 \\
       0 & -1 &  0 &  0 \\
       0 &  0 & -1 &  0 \\
       0 &  0 &  0 & -1 \\
     \end{array}
   \right),
\end{equation}
\begin{equation}\label{eq69}
T_{32} =\left(
     \begin{array}{cccc}
        1 & 0 &            0                           &        0           \\
        0 & 1 &            0                           &        0           \\
        0 & 0 & -\frac{\sqrt{2}}{2} & \frac{\sqrt{2}}{2} \\
        0 & 0 & -\frac{\sqrt{2}}{2} & -\frac{\sqrt{2}}{2} \\
     \end{array}
   \right),
\end{equation}
\begin{equation}\label{eq70}
T_{31} =\left(
     \begin{array}{cccc}
        1 &         0                            & 0 &         0                             \\
        0 & -\frac{\sqrt{6}}{3} & 0 & \frac{\sqrt{3}}{3} \\
        0 &         0                            & 1 &          0                             \\
        0 & -\frac{\sqrt{3}}{3} & 0 & -\frac{\sqrt{6}}{3} \\
     \end{array}
   \right),
\end{equation}
\begin{equation}\label{eq71}
T_{30} =\left(
     \begin{array}{cccc}
       \frac{1}{2}                & 0 & 0 & \frac{\sqrt{3}}{2} \\
            0                                & 1 & 0 &        0                             \\
            0                                & 0 & 1 &        0                             \\
        -\frac{\sqrt{3}}{2} & 0 & 0 &  \frac{1}{2}       \\
     \end{array}
   \right),
\end{equation}
\begin{equation}\label{eq72}
T_{21} =\left(
     \begin{array}{cccc}
        1 &         0                           &        0                             & 0 \\
        0 & -\frac{\sqrt{3}}{2} & \frac{1}{2}                & 0 \\
        0 & -\frac{1}{2}              & -\frac{\sqrt{3}}{2} & 0 \\
        0 &         0                              &         0                             & 1 \\
     \end{array}
   \right),
\end{equation}
\begin{equation}\label{eq73}
T_{20} =\left(
     \begin{array}{cccc}
       \frac{\sqrt{3}}{3}   & 0 & \frac{\sqrt{6}}{3} & 0 \\
               0                            & 1 &         0                            & 0 \\
        -\frac{\sqrt{6}}{3} & 0 & \frac{\sqrt{3}}{3} & 0 \\
               0                             & 0 &         0                            & 1 \\
     \end{array}
   \right),
\end{equation}
\begin{equation}\label{eq74}
T_{10} =\left(
     \begin{array}{cccc}
         \frac{\sqrt{2}}{2} & \frac{\sqrt{2}}{2} & 0 & 0 \\
        -\frac{\sqrt{2}}{2} & \frac{\sqrt{2}}{2} & 0 & 0 \\
                0                           &        0                             & 1 & 0 \\
                0                          &        0                              & 0 & 1 \\
     \end{array}
   \right).
\end{equation}
The correspondences between $\mathcal{T}_4$ with $c_0=c_1=c_2=c_3=\frac{1}{2}$, parameters $\theta_{ij}$ and $\phi_{ij}$ are shown in Tab. \ref{table1}.

\begin{table}[h!]
  \centering
  \caption{The correspondences between $\mathcal{T}_4$ with $c_0=c_1=c_2=c_3=1/2$, $\theta_{ij}$, and $\phi_{ij}$.}
  \label{table1}
  \begin{tabular}{ccc}
      \hline \hline
\quad $T_{ij}(\theta_{ij},\phi_{ij})$ \qquad & \qquad $ \theta_{ij}$        \qquad & \qquad $\phi_{ij}$ \quad \\
      \hline
      $T_{32}$                        \qquad & \qquad $ \pi/4$              \qquad & \qquad $\pi$  \\
      $T_{31}$                        \qquad & \qquad $ \arctan(\sqrt{2})$  \qquad & \qquad $\pi$  \\
      $T_{30}$                        \qquad & \qquad $ \pi/3$              \qquad & \qquad $0$   \\
      $T_{21}$                        \qquad & \qquad $ \pi/3$              \qquad & \qquad $\pi$ \\
      $T_{20}$                        \qquad & \qquad $ \arctan(\sqrt{2})$  \qquad & \qquad $0$   \\
      $T_{10}$                        \qquad & \qquad $ \pi/4$              \qquad & \qquad $0$   \\
      \hline \hline
  \end{tabular}
\end{table}

\subsection{The evaluations of the collective operations $W$ and $Z$}\label{sec43}

It is known that some high-dimensional single-partite gates, such Pauli $X$ and SUM gates, in OAM have been experimental demonstrated  \cite{single-qudit}.
However, realizations of multi-partite high-dimensional gates are usually challenging both in theory and experiment \cite{two-qudit1,two-qudit2,two-qudit3}.
Fortunately, in our RSP schemes via non-maximally entangled states, the two-qubit collective operations $W$ given in Eq. \eqref{eq20} and $Z$  given in Eq. \eqref{eq57} can be avoided by encoding the computational basis in the spatial DOFs of single-photon systems.

\begin{figure}[htbp]
\centering
\includegraphics[width=5 cm]{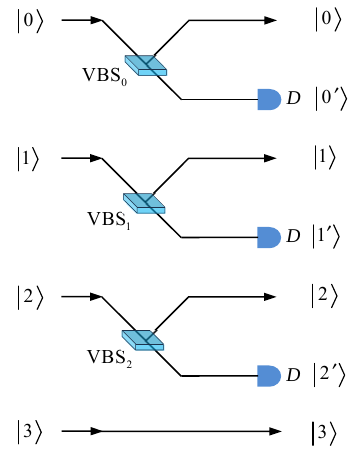}
\caption{Quantum circuit for  {concentrating
$|\Psi\rangle_{AB} = \frac{1}{2}(|00\rangle + |11\rangle + |22\rangle + |33\rangle)_{AB}$
from
$|\tilde{\Psi}\rangle_{AB} = (a_0|00\rangle + a_1|11\rangle + a_2|22\rangle + a_3|33\rangle)_{AB}$.} VBS is a variable reflectivity beam splitter and it can be completed by Fig. \ref{Figure1}(b). $D$ is a single-photon detector.} \label{Figure2}
\end{figure}

As shown in Fig. \ref{Figure2}, if the coefficients $a_0$, $a_1$, $a_2$, and $a_3$ are known to Bob,
the spatial-based partially entangled state $|\tilde{\Psi}\rangle_{AB} = (a_0|00\rangle + a_1|11\rangle + a_2|22\rangle + a_3|33\rangle)_{AB}$ given in Eq. \eqref{eq14}
can be concentrated into the maximally entangled state $|\Psi\rangle_{AB} = \frac{1}{2}(|00\rangle + |11\rangle + |22\rangle + |33\rangle)_{AB}$ given in Eq. \eqref{eq2}
by setting three variable reflectivity beam splitters VBS$_0$, VBS$_1$, and VBS$_2$ on spatial modes 0, 1, and 2 of photon B, respectively.
Here reflection coefficients of VBS$_0$, VBS$_1$, and VBS$_2$ are adjusted to $\frac{a_3}{a_0}$, $\frac{a_3}{a_1}$, and $\frac{a_3}{a_2}$, respectively.
That is, the three VBSs accomplish the following transformations
\begin{equation}\label{eq75}
\begin{split}
& a_0|0\rangle \xrightarrow{\text{VBS}_0} a_3|0\rangle  + \sqrt{(a_0)^2 - (a_3)^2} |0'\rangle, \\
& a_1|1\rangle \xrightarrow{\text{VBS}_1} a_3|1\rangle  + \sqrt{(a_1)^2 - (a_3)^2} |1'\rangle, \\
& a_2|2\rangle \xrightarrow{\text{VBS}_2} a_3|2\rangle  + \sqrt{(a_2)^2 - (a_3)^2} |2'\rangle.
\end{split}
\end{equation}
Therefore, three VBSs transform $|\tilde{\Psi}\rangle_{AB}$  into
\begin{equation}\label{eq76}
\begin{split}
|\overline{\Psi}_{AB}\rangle_{AB} =\; & a_3 (|00\rangle + |11\rangle  + |22\rangle   + |33\rangle)_{AB}  \\ &+
                                             \sqrt{(a_0)^2 - (a_3)^2}|00'\rangle_{AB} \\ & +
                                             \sqrt{(a_1)^2 - (a_3)^2}|11'\rangle_{AB} \\ & +
                                             \sqrt{(a_2)^2 - (a_3)^2}|22'\rangle_{AB}.
\end{split}
\end{equation}
Based on Eq. \eqref{eq76}, we can see that the click of any one detector means the scheme  fails. Otherwise, the system composed of particles $A$ and $B$ will collapse to the normalized state $|\Psi_0\rangle_{AB}$ with a success probability of $4|a_3|^2$.

A similar arrangement as that made in Fig. \ref{Figure2},
$|\tilde{\Phi}\rangle_{AB} = (b_0|00\rangle + b_1|11\rangle + b_2|22\rangle + b_3|33\rangle + b_4|44\rangle + b_5|55\rangle + b_6|66\rangle + b_7|77\rangle)_{AB}$
described by Eq. \eqref{eq47}
can be concentrated into
$|\Phi\rangle_{AB}= \frac{1}{2\sqrt{2}}(|00\rangle + |11\rangle + |22\rangle + |33\rangle +|44\rangle +|55\rangle + |66\rangle + |77\rangle)_{AB}$
described by Eq. \eqref{eq26}
by setting VBS$_0$, VBS$_1$, $\cdots$, VBS$_6$, with reflection coefficients
$\frac{b_7}{b_0}$, $\frac{b_7}{b_1}$, $\cdots$, $\frac{b_7}{b_6}$ on spatial modes 0, 1, $\cdots$, 6 of photon B, respectively.

\section{Discussions and conclusions}\label{sec5}

In our proposed schemes,  RSPs of single-partite state in four- (eight-) dimensional complex Hilbert space and real subspace are designed, respectively. Moreover, in our RSPs,  both the maximally and the non-maximally quantum channels are taken into accounted.
It is worthy to point out that, as shown in Tab. \ref{Table1}, RSP in three-level system is the special case of Eq. \eqref{eq1}, that is, $c_3=0$.
Similarly, this also happens in the case of Eq. \eqref{eq25}, five-, six-, and seven-level RSPs corresponding to $d_{5}=d_{6}=d_{7}=0$, $d_{6}=d_{7}=0$, and $d_7=0$, respectively.

\begin{table}[htb]
\centering \caption{The special case of the present high-dimensional RSP schemes. We take $\theta_0=0$ and $\vartheta_i=0$.}

\begin{tabular}{ccccccccc}

\hline  \hline

    multi-level  state                          \quad & \quad 3-level  \quad & \quad 5-level \quad & \quad  6-level \quad & \quad 7-level \\  \hline

$\sum_{i=0}^4 c_i e^{\text{i}\theta_i} |i\rangle$     &      $c_3=0$         &                     &                      &            \\

$\sum_{i=0}^7 d_i e^{\text{i}\vartheta_i} |i\rangle$  &                      &  $d_{5,6,7}=0$      & $d_{6,7}=0$          &  $d_7=0$     \\
                             \hline  \hline
\end{tabular}\label{Table1}
\end{table}

By encoding the qudit and qunit on the spatial modes of single-photon systems, our high-dimension RSPs can be accomplished by using linear optical elements, including BS, VBS, and PS, see Sec. \ref{sec42} and Sec. \ref{sec43}.
However, in practical applications, the imperfect BS and PS will degrade the fidelities of our RSPs.
As shown in Fig. \ref{Figure1}(b), the transformation matrices of the two imperfect 50:50 BSs can be given by \cite{imperfection}
\begin{equation}\label{eq77}
\tilde{U}_{\text{BS}}=\frac{1}{\sqrt{{\epsilon}^{2}+2\epsilon+2}}
        \left(
           \begin{array}{cc}
            1 + \epsilon &  \text{i}   \\
	           \text{i}          &  1+\epsilon \\
          \end{array}
       \right),
\end{equation}
where $\epsilon$ is the imperfection of the transmission ratio, and it deviates slightly from 50\%. The unitary transformation matrices of imperfect PS$(2\theta)$ and PS$(\phi)$ can be given by  \cite{imperfection}
\begin{equation}\label{eq78}
\tilde{U}_{\text{PS}(2\theta)}=
        \left(
           \begin{array}{cc}
            e^{\text{i}2(\theta-\delta\theta)} &  0  \\
	                      0                                                   &  1  \\
          \end{array}
       \right),
\end{equation}
\begin{equation}\label{eq79}
\tilde{U}_{\text{PS}(\phi)}=
        \left(
           \begin{array}{cc}
            e^{\text{i}(\phi-\delta\phi)} &  0  \\
	                     0                                         &  1  \\
          \end{array}
       \right),
\end{equation}
where $(\theta-\delta\theta)$ and $(\phi-\delta\phi)$ are the real phases difference between the two arms of the MZI.
Hence, the transformation of block $T(\phi,\theta)$ becomes
\begin{equation}\label{eq80}
  \begin{split}
&\tilde{T}(\phi,\theta)
 = \frac{1}{(1+\epsilon)^2+1} \\
 &\begin{pmatrix} \left((1+\epsilon)^2 e^{\text{i}2(\theta-\delta\theta)} - 1\right) e^{\text{i}(\phi-\delta\phi)}
   & \text{i}(1+\epsilon)\left(e^{\text{i}2(\theta-\delta\theta)} + 1\right) \\
  \text{i}(1+\epsilon)\left(e^{\text{i}2(\theta-\delta\theta)} + 1\right) e^{\text{i}(\phi-\delta\phi)}
  & (1+\epsilon)^2 - e^{\text{i}2(\theta-\delta\theta)} \end{pmatrix}
 \end{split}
\end{equation}%

Without any loss of generality, the input state of the $\tilde{T}(\phi,\theta)$ is prepared as
\begin{equation}\label{eq81}
|\chi_{\text{in}}\rangle = \cos \alpha |0\rangle + \sin \alpha|1\rangle.
\end{equation}

Based on Eq. \eqref{eq66}, in the ideal case, the output state of $T(\phi,\theta)$ can be calculated as
\begin{equation}\label{eq82}
\begin{split}
&|\chi_{\text{out}}\rangle_{\text{idea}} \\
&= \left( \frac{1}{2} \left[ (e^{\text{i}2\theta} - 1)e^{\text{i}\phi} \cos \alpha  + \text{i}(e^{\text{i}2\theta} + 1) \sin \alpha \right] \right)|0\rangle \\
&+ \left( \frac{1}{2} \left[ \text{i}(e^{\text{i}2\theta} + 1)e^{\text{i}\phi} \cos \alpha  + (1 - e^{\text{i}2\theta}) \sin \alpha \right] \right)|1\rangle.
\end{split}
\end{equation}
Based on Eq. \eqref{eq80}, the realistic case, the output state of $T(\phi,\theta)$ can be calculated as
\begin{equation}\label{eq83}
\begin{split}
&|\chi_{\text{out}}\rangle_{\text{real}}\\
&= \left( \frac{1}{(1+\epsilon)^2+1} \left[ \left((1+\epsilon)^2 e^{\text{i}2(\theta-\delta\theta)} - 1\right) e^{\text{i}(\phi-\delta\phi)} \cos \alpha \right. \right. \\
&\left. \left. + \text{i}(1+\epsilon)\left(e^{\text{i}2(\theta-\delta\theta)} + 1\right) \sin \alpha \right] \right)|0\rangle \\
&+ \left( \frac{1}{(1+\epsilon)^2+1} \left[ \text{i}(1+\epsilon)\left(e^{\text{i}2(\theta-\delta\theta)} + 1\right) e^{\text{i}(\phi-\delta\phi)} \cos \alpha \right. \right. \\
& \left. \left. + \left((1+\epsilon)^2 - e^{\text{i}2(\theta-\delta\theta)}\right) \sin \alpha \right] \right)|1\rangle.
\end{split}
\end{equation}

\begin{figure}[htbp]
\centering
\includegraphics[width=8 cm]{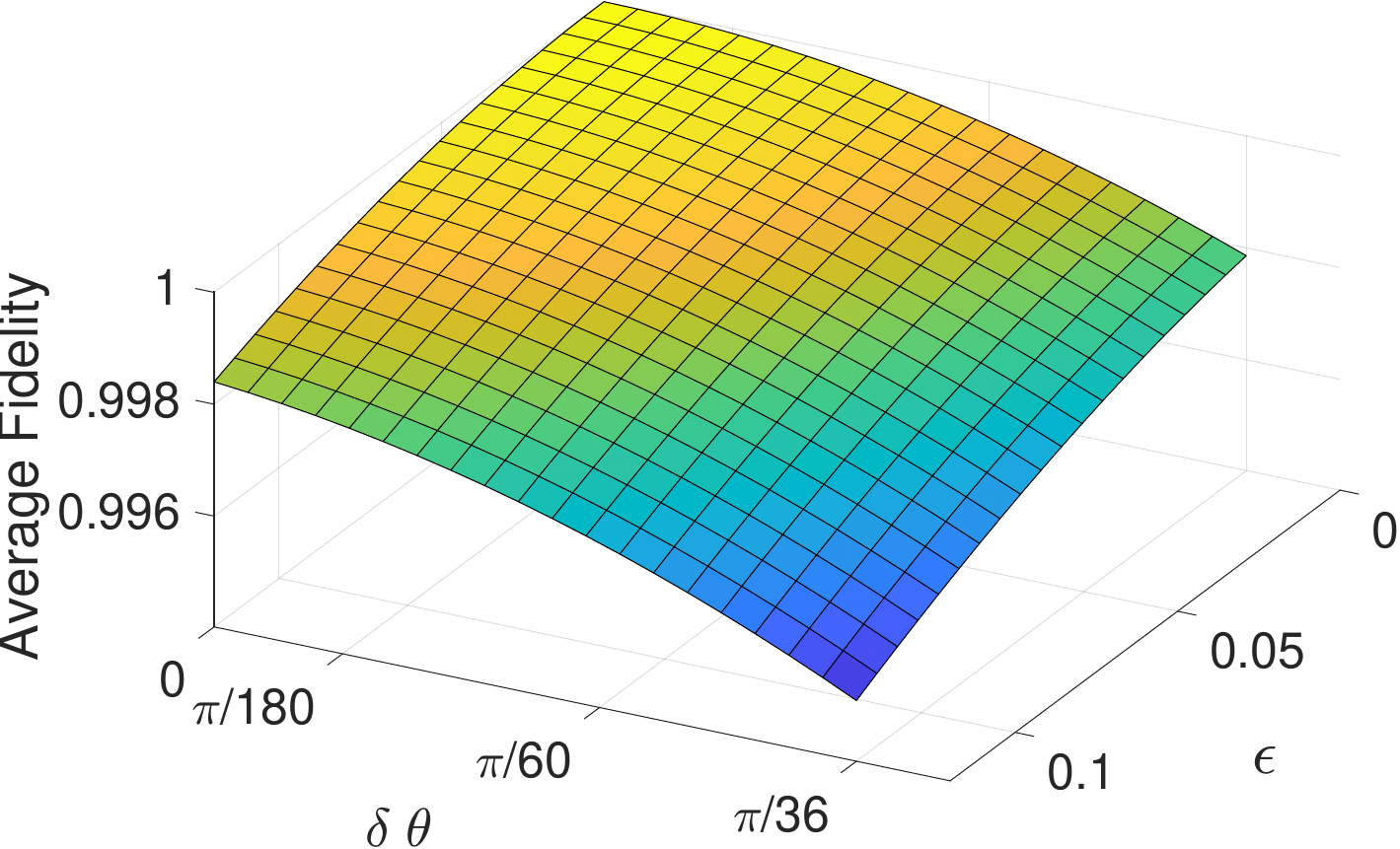}
\caption{Average fidelity of $\tilde{T}(\phi,\theta)$. $\delta\theta=\delta\phi$, $\theta=\pi/3$, and $\phi=0$ are taken.} \label{Figure3}
\end{figure}

The average fidelity $ \bar{F} = \frac{1}{2\pi} \int_0^{2\pi} |{}_{\text{idea}}\langle \chi_{\text{out}}| \chi_{\text{out}}\rangle_{\text{real}}|^2 \text{d} \alpha $ of $T(\phi,\theta)$ is shown in Fig. \ref{Figure3}.
Here $\delta\theta=\delta\phi$, $\theta=\pi/3$, and $\phi=0$ are taken.
The efficiency of $T(\phi,\theta)$ is unit in principle.
The dark counts and dark counts on single-photon detections  also reduce the performance of $T(\phi,\theta)$ (5\%-10\%).

In conclusion, we have proposed various remote equatorial state preparation for four- and eight-dimensional systems.
The previous qubit-based RSPs mainly focused on  {real subspace within the complex Hilbert space} using maximally entangled pair \cite{Pati,Lo}.
Yu \emph{et al}. \cite{qudit-Song} proposed a RSP scheme, and the qudit is prepared onto a group of particles of qubit in {real subspace within the complex Hilbert space}.
The RSPs we proposed allow for four-level and eight-level equatorial states in both the complex and real Hilbert spaces,
and both the pre-shared maximally and partially entangled states are taken into account.
A set of orthogonal measurement basis is identified to implement our schemes, and this projection measurement is relatively easier to be experimentally realized  compared to the complex POVM measurements. Note that the RSP of three-level (five-, six-, and seven-level) equatorial state can also be obtained from our scheme for the four-level (eight-level) RSP.
The evaluation of the schemes indicates that our scheme might be possible with current technology. The collection operations, necessary for partially-entangled-based RSPs, can be avoided by encoding the qudit on the spatial modes of single-photon systems.  The spatial-based single-qudit operation can be achieved by using some  linear optical elements \cite{BS1994,BS2016,UBS3,UBS4,multiparty optical}.

\section*{Acknowledgments}\label{sec5}
This work is supported by the National Natural Science Foundation of China under Grant No. 62371038, the Science Research Project of Hebei Education Department under Grant No. QN2025054, and the Fundamental Research Funds for the Central Universities under Grant No. FRF-TP-19-011A3.
\medskip

\vspace{6 cm}



\end{document}